\documentclass[10pt,conference]{IEEEtran}
\usepackage{cite}
\usepackage{amsmath,amssymb,amsfonts}
\usepackage{xcolor}
\usepackage{hyperref}
\usepackage{listings}
\usepackage{tabularx}
\usepackage{microtype}
\usepackage{tcolorbox}
\usepackage{bm}
\usepackage{amsmath}
\usepackage{makecell}
\usepackage[vlined,ruled,linesnumbered]{algorithm2e}
\usepackage{multirow}
\usepackage{makecell}
\usepackage{bold-extra}
\usepackage[T1]{fontenc}
\usepackage{enumitem}

\usepackage{mathtools}
\usepackage{tikz}
\usepackage{forest}

\usepackage[font=small,labelfont=bf,tableposition=top]{caption}

\DeclareCaptionLabelFormat{andtable}{#1~#2  \&  \tablename~\thetable}

\setlist[itemize]{noitemsep, topsep=0pt}

\colorlet{defgreen}{green!35!black}

\tcbuselibrary{theorems}
\newtcbtheorem[number within=section]{cdefinition}{Definition}%
{colback=green!5,colframe=defgreen,colbacktitle=green!40!black!30,coltitle=black,fonttitle=\bfseries,boxsep=1pt,left=2pt,right=2pt,top=0pt,bottom=0pt}{th}

\def\oracle{\ensuremath{\phi{}}}

\usepackage{cleveref}
\usepackage{marginnote}
\usepackage{adjustbox}

\usepackage{booktabs} 
\usepackage{amssymb}
\usepackage{pifont}
\usepackage{multirow}

\usepackage{extarrows}

\usepackage{graphicx,fancybox}
\usepackage{mathtools}
\usepackage{listings}
\usepackage{url}
\usepackage{tcolorbox}
\usepackage{microtype}
\usepackage{hyperref}
\usepackage{xcolor}
\usepackage{color, colortbl}

\usepackage{array}
\usepackage{lipsum}
\newcolumntype{P}[1]{>{\centering\arraybackslash}p{#1}}
\newcolumntype{L}[1]{>{\raggedright\arraybackslash}p{#1}}
\usepackage[para]{threeparttable}

\usepackage{textcomp}
\usepackage{pgfplots}

\pgfplotsset{compat=1.10}
\usetikzlibrary{decorations.pathreplacing,calc,trees,arrows,shapes,decorations,positioning}
\usepackage{lmodern}

\SetEndCharOfAlgoLine{}
\SetAlFnt{\sffamily\footnotesize}
\SetKwProg{Fn}{Function}{}{}
\SetKwBlock{Loop}{Loop}{EndLoop}

\SetFuncArgSty{\sffamily{}}

\definecolor{arboristgreen}{rgb}{0, 0.35, 0}

\tikzset{
  branch/.style={
    color=blue!60!black,
    thick
  },
  leaf/.style = {
    text=black, shape=rectangle, draw=arboristgreen, thick, align=center, top color=white, bottom color=blue!80!black!40,
    inner sep=2pt
  },
  internal/.style = {
    text=black, shape=rectangle, rounded corners, draw=arboristgreen, thin, align=center, top color=white, bottom color=blue!60!yellow!50,
    inner sep=2pt
  },
  buggy/.style = {
    draw=red, dashed, ultra thick
  }
}

\tcbuselibrary{theorems}

\def\algosize{\small}

\def\oracle{\ensuremath{\phi{}}}

\usepackage{fullpage}
\usepackage{calc}
\usetikzlibrary{positioning,shadows,arrows,trees,shapes,fit}
\usepackage{color,soul}
\usepackage{subcaption}

\usepackage{ifthen}
\usepackage{pstricks-add}
\usepackage{siunitx}

\begin{document}

\title{Leveraging Models to Reduce Test Cases in Software Repositories}

\author{\IEEEauthorblockN{Golnaz Gharachorlu}
\IEEEauthorblockA{\textit{Simon Fraser University, Canada}\\
ggharach@sfu.ca}
\and
\IEEEauthorblockN{Nick Sumner}
\IEEEauthorblockA{\textit{Simon Fraser University, Canada}\\
wsumner@sfu.ca}
}

\maketitle

\begin{abstract}
Given a failing test case,
test case reduction yields a smaller test case that reproduces the failure.
This process can be time consuming
due to repeated trial and error with smaller test cases.
Current techniques speed up reduction by only
exploring syntactically valid candidates,
but they still spend significant effort on
semantically invalid candidates.
In this paper, we
propose a model-guided approach to speed up
test case reduction.
The approach trains a model of semantic properties
driven by syntactic test case properties. 
By using this model, we can skip 
testing even syntactically valid test case candidates that are
unlikely to succeed. 
We evaluate this model-guided reduction on a suite of 14 large fuzzer-generated C
test cases from the bug repositories of two well-known C compilers, GCC and Clang. Our results show
that with an average precision of 77\%, we can
decrease the number of removal trials by 14\% to
61\%. 
We observe a 30\% geomean improvement in reduction time over the
state of the art technique while preserving similar
reduction power.
\end{abstract}


\section{Introduction}
\label{intro}
In many software issue trackers, only bugs that have a minimal reproducible test case are investigated~\cite{elm-reduce}. Manually reducing test cases is difficult and time-consuming~\cite{DBLP:conf/pldi/RegehrCCEEY12}. Hence, an automated test case reduction technique is necessary to overcome the barriers to reporting bugs and facilitate testing and debugging programs. Given a test case that demonstrates a bug in a program, the goal of test case reduction is to achieve a smaller test case that exhibits the same bug in the program but is easier to understand~\cite{DBLP:journals/tse/ZellerH02, DBLP:conf/icse/Misherghi06,DBLP:conf/icse/SunLZGS18, DBLP:conf/fase/GharachorluS19}.
The reduction continuously performs \emph{removal trials} by pruning parts of the test case
and running an oracle, usually using the program with the bug, on the resulting smaller test case to determine whether the bug still occurs.
When the oracle returns true, the original test case is replaced with the reduced one and reduction continues to find smaller test cases.
The process terminates when no further removals succeed.

There is a variety of state of the art reduction techniques. Among them are the well-known Delta Debugging~\cite{DBLP:journals/tse/ZellerH02} and Hierarchical Delta Debugging~\cite{DBLP:conf/icse/Misherghi06}, the language specific reducer C-Reduce~\cite{DBLP:conf/pldi/RegehrCCEEY12} and more recently, two queue-driven approaches called Perses~\cite{DBLP:conf/icse/SunLZGS18} and Pardis~\cite{DBLP:conf/fase/GharachorluS19}.
The two last techniques leverage a priority queue to traverse nodes in the abstract syntax tree (AST) of a test case and try removing subtrees of the AST rooted at these nodes.
Both these techniques only explore nodes that are syntactically removable.
For instance, elements of a list may be removed while preserving the syntactic validity of the entire list.
However, a node representing a function name within a declaration cannot be removed,
as its removal would result in a declaration with no name.
Using a more efficient priority mechanism, Pardis outperforms Perses in terms of reduction time and number of tests\footnote{
In this paper, we use the terms oracle queries and tests interchangeably to refer to the pruning (removal) trials.}.
By avoiding syntactically invalid inputs, these approaches guide the search toward smaller test cases that pass initial input parsing.

Although \emph{syntactic} validity is a major factor when removing nodes, reduction may still explore invalid inputs.
In reducing test cases, many nodes are syntactically valid to remove but will lead to test cases that are not \emph{semantically} valid.
For instance, removing the declaration of a variable before removing its use causes a compiler to complain about the \textit{use of undeclared identifier}.
In the latest reduction techniques, these removal trials are still performed, and the oracle is run to verify whether the bug is preserved.
However, these reduced test cases are not semantically valid, and the oracle will return false.
Hence, it is desirable to avoid running these tests considering that querying an oracle is often an expensive task \cite{oracle-expensive}. 

In this paper, we address the problem of semantically invalid tests and propose a technique to mitigate it by training models that are able to predict whether removing a node from the AST can cause semantic invalidity. Given the queue of syntactically removable nodes, we query our model before querying the oracle in our attempt to remove each node in the queue. The model has two possible outcome suggestions:

1) Removing the node \textit{does not generate} a semantically invalid test case: We continue with querying the oracle to see if the node can be removed.

2) Removing the node \textit{generates} a semantically invalid test case: We skip the trial, avoid running the oracle, and proceed with the next node in the queue.

We need to have models that can correctly skip semantically invalid tests to improve the \textit{performance} of the reduction. However, they should also have a high \textit{precision} and be able to consider as many tests that are semantically valid as possible without skipping them in order to achieve reduced test cases with size similar to state of the art techniques.

To this end, we make the following contributions:
\begin{itemize}
    \item We address the problem of semantic invalidity in test case reduction. Using random forests \cite{DBLP:conf/icica/LiuWZ12}, we train models capable of predicting whether removing a node from the AST of the test case is semantically valid. We leverage different syntactic test case properties as features for predicting semantic properties.
    We use these features both individually and in combination when training our models.
    \item We evaluate our models on a suite of 14 large fuzzer-generated test cases in C, extracted from GCC and Clang bug report repositories and also used in Perses and Pardis studies. We introduce our best model in terms of the final test case reduction time and the number of oracle queries.
    \item We measure precision and recall rates of our models when reducing our benchmark in addition to measuring them on a different set of test data.
    \item We conduct a study to reason about true negatives and false positives of our models with respect to the different types of semantic issues.
\end{itemize}

Compared to Pardis, we are able to reduce the number of oracle queries by 14\% to 61\% on our 14 test cases leading to an improvement of 30\% (geomean) in the reduction time.
Compared to C-Reduce, we decrease the reduction time by an average of 57\%.
The precision and recall rates of our three main models are above 70\% and 77\% on average.
In addition to comparing our models against Pardis and C-Reduce, we use them as a \emph{preprocessing step} on these techniques and improve their performance \emph{while generating reduced test cases of the same size}.

The rest of the paper is organized as follows: In the next section, we provide a background on the state of the art syntax-guided reduction techniques and describe semantic invalidity. 
In~\autoref{approach}, we explain our approach including data collection, feature extraction, training and querying the models. \Cref{eval} provides the performance results of our approach along with the precision and recall rates of the models and how they behave with respect to each type of semantic issue.
We discuss possible improvements and future directions in \autoref{discuss}. 
\section{Background and Motivation}
\label{motiv}
To reduce a test case with a property of interest, such as exhibiting a bug, state of the art approaches parse an input into an abstract syntax tree (AST) of the test case, and each node of the AST is considered for removal.
An oracle function, $\psi:\mathbb{I}\rightarrow\mathbb{B}$ determines whether the removal of the node is successful. This function returns true iff the property of interest still holds after removing the node (the same bug still occurs); i.e., given $T$ as the AST of the test case to reduce, removal of node $n$ is successful iff $T-n$ generates an input $I\in\mathbb{I}$ such that $\psi\left(I\right)=\mbox{true}$. 
An oracle function typically checks multiple aspects of a candidate test case \textit{before} passing the input to the original program and verifying whether it holds the property of interest.
These aspects along with the verification of the bug make running such oracles an expensive task, especially for complex test cases and inputs with a long running time.

We describe four steps necessary for an oracle to take before generating the Boolean outcome:
\begin{enumerate}
    \item \textbf{Syntactic check:} Removing the node does not violate syntactic rules of the test case's grammar.
    \item \textbf{Semantic check:} Removing the node does not violate semantic rules of the test case's grammar.
    \item \textbf{Well-definedness check:} Removing the node does not lead to undefined or indeterminate behavior.
    \item \textbf{Property of interest check:} Removing the node produces a test case variant that exhibits the same property of interest.
\end{enumerate}

Although steps one and four of the oracle checks have been intensively addressed in the literature, steps two and three have drawn less attention.
Initial techniques such as Delta Debugging (DD)~\cite{DBLP:journals/tse/ZellerH02} and Hierarchical Delta Debugging (HDD)~\cite{DBLP:conf/icse/Misherghi06} do not consider the first three steps. By default, they run tests to check the property of interest on a large number of variants that may be meaningless. State of the art techniques such as Perses~\cite{DBLP:conf/icse/SunLZGS18} and Pardis~\cite{DBLP:conf/fase/GharachorluS19} leverage step one to improve the efficiency reduction by pruning the search space to only consider nodes that are syntactically removable.
Thus, Perses and Pardis consider a smaller search space for inputs ($P_1$) than that of DD and HDD ($P_0$), or simply $P_1\subseteq P_0$.

\autoref{fig::nullable} shows part of a context free grammar with the rule \texttt{C\_opt} tagged with quantifier \texttt{?}. This quantifier indicates that the rule type is optional; i.e., a node with this type can safely get removed from the AST without causing any syntactic issues. Hence, the program generated by removing this node is within $P_1$.  However, a syntactically removable node can be semantically invalid to remove. 

\begin{figure}
\centering
\begin{minipage}{.43\textwidth}
\scriptsize
\centering
\begin{tabular}{|rcl|}
\hline
\sffamily
C $\rightarrow$ A $|$ A B
&$\Longrightarrow$&
\sffamily
\makecell[l{m{0.33\linewidth}}]{
C $\rightarrow$ A C\_opt\\
\sffamily
C\_opt $\rightarrow$ B?
}\\
\hline
\end{tabular}
\vspace{.5em}
\caption{Removing C\_opt is syntactically valid but may not be semantically.}
\label{fig::nullable}
\end{minipage}%
\hspace{1em}%
\begin{minipage}{.55\textwidth}
  \centering
  \input{texFigures/motivating-example-Listing}
 \end{minipage}
\end{figure}
\tikzset{nullable/.append style={draw=red, ultra thick}}
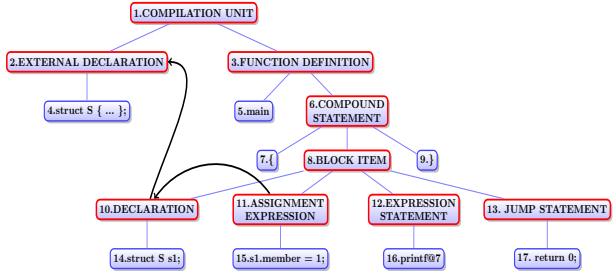
\begin {figure}
\centering
\begin{adjustbox}{width=\columnwidth, height = 1.4 in}
\begin{tikzpicture}
[font=\normalsize, edge from parent, 
    every node/.style={top color=white, bottom color=blue!25, 
    rectangle,rounded corners, minimum size=6mm, draw=blue!75,
    very thick, drop shadow, align=center,font=\bfseries},
    edge from parent/.style={draw=blue!50,thick},
    level 1/.style={sibling distance=8cm},
    level 2/.style={sibling distance=3.5cm}, 
    level 3/.style={sibling distance=3cm}, 
    level 4/.style={sibling distance=5cm}, 
    level 5/.style={sibling distance=8cm}, 
    level distance=1.5cm,
    ]
    \node [nullable](A){1.COMPILATION UNIT} 
        child { node [nullable] (B) {2.EXTERNAL DECLARATION}
            child { node {4.struct S \{ ... \};}
            }
            }
    child { node [nullable] {3.FUNCTION DEFINITION} 
        child { node {5.main}}
        child { node [nullable] {6.COMPOUND\\ STATEMENT}
               child {node {{7.\{}}}
               child {node [nullable] {8.BLOCK ITEM}
                    child {node [nullable](C){10.DECLARATION}
                            child {node {14.struct S s1;}
                            }
                    }
                   child {node [nullable](D){11.ASSIGNMENT\\ EXPRESSION}
                        child {node {15.s1.member = 1;}}
                   }
                   child {node[nullable] {12.EXPRESSION\\  STATEMENT}
                       child {node {16.printf@7}}
                   }
                   child {node[nullable] {13. JUMP STATEMENT}
                        child {node {17. return 0;}}
                   }
               }
               child {node {{9.\}}}}
        }
};
\draw [->,ultra thick ] (C) to [out=70,in=0] (B);
\draw [->,ultra thick ] (D) to [out=130,in=50] (C);

\end{tikzpicture}
\end{adjustbox}
\caption{AST of \autoref{listing::motivatingExample}. Red frames depict syntactically removable nodes. Removing node 2 before 10 and 11 or removing 10 before 11 causes semantic invalidity.}
\label{motiv-fig}
\vspace{-1em}
\end{figure}
 For instance, in one C grammar~\cite{antlr-c}, it is syntactically valid to remove a \texttt{declaration} of a variable or \texttt{definition} of a function, but if their uses are still in the test case, they cannot get removed without causing the compiler to complain about undeclared or undefined entities. \autoref{motiv-fig} depicts the simplified AST of the program in \autoref{listing::motivatingExample}. In this example, node 11 is dependent on node 10 which itself is dependent on node 2. Although all three nodes are tagged as syntactically removable, they need to get removed in an appropriate order to avoid semantic validity issues.

In this paper, we prune the search space even more by leveraging models to predict semantic validity of a node removal. Our models are trained by extracting syntactic properties of our large program corpus described in \autoref{approach}. Our search space, $P_2\subseteq P1$ consists of programs that are predicted as semantically valid by our models. For instance, a test case with a declaration and use of a variable could be in $P_2$ while a test case with only the use of a variable (declaration removed) is not. We run oracles on test cases that belong to $P_2$ and discard the ones that do not.

We define a test as semantically invalid if it causes at least one of the following:
\begin{itemize}
    \item \textbf{Compilation errors:} Any test case variant that cannot get compiled successfully and causes compiler to return failure exit status.
    \item \textbf{Blacklisted warnings:} Any test case variant that generates a warning message that is in the list of blacklisted warnings extracted from the oracle scripts in Perses \cite{oracle-perses} and Pardis \cite{oracle-pardis}. The list includes a range of warnings from \textit{missing type specifier} and \textit{empty declaration} to \textit{incompatible conversions}. 
\end{itemize}

\section{Approach}
\label{approach}
We propose an approach to leverage models to predict whether removal of a node from the AST of a test case is semantically valid or not. In other words, does removal of a node generate a test case that causes compilation errors or blacklisted warnings described in \autoref{motiv}? We describe the process of collecting data and extracting features for training our predictive models. In this paper, we focus on reducing C programs as test cases for compilers.
However, our approach is general and could apply to any input domain with a grammar.

\subsection{Training Data Collection}
\label{data-collection}
To train models capable of predicting semantic validity, we need to collect training data in form of \texttt{(source, target)} pairs such that \texttt{source} contains the features representing the node and \texttt{target} is a Boolean value that shows a valid or invalid removal. To this end, we collect our set of training data using the GCC torture test suite\footnote{https://github.com/gcc-mirror/gcc/tree/master/gcc\\/testsuite/gcc.c-torture}.
This suite consists of a variety of C test cases with different ASTs that enable us to obtain a comprehensive dataset. These test cases are designed to intensely examine GCC and cover a wide range of the compiler's behavior. In addition, the relatively small size of these test cases allows us to process and collect data from each file efficiently. 
We randomly select 600 test cases from this suite. For each test case, we build its AST and pass it to Pardis \cite{DBLP:conf/fase/GharachorluS19}, the latest state of the art syntax-based program reducer. As described in \autoref{intro} and \autoref{motiv}, Pardis places only syntactically removable nodes in a priority queue and removes them one at a time from the AST. Hence, by eliminating the possibility of removing syntactically invalid nodes, Pardis enables us to capture the removal success or failure of nodes with respect to their semantic validity. We define an oracle function and pass it to Pardis as a script along with the AST of the test case.
The oracle returns true when both of the following criteria are satisfied:

\begin{itemize}
\item{Removing the node from the AST does not yield semantic invalidity.}
\item{Removing the node from the AST does not remove a specific randomly selected token from the test case.}
\end{itemize}

The second criterion prevents early termination of a reduction phase and enables us to collect more data.

\autoref{app-fig} depicts the process of collecting and logging ASTs from test cases. By trying to remove each node, we capture an AST in which the node considered for removal (the \textit{query node}), is tagged as the source along with a success (\textit{s}) or failure  (\textit{f}) target value that is the outcome of the oracle on the candidate when the query node is removed.

If the query node \textit{cannot be removed}, we record the AST, the node and the failure outcome and continue with trying to remove the next node in the queue.

If the query node \textit{can be removed} successfully, we record the AST, the node and the success outcome. Then we remove the node from the AST and continue with trying to remove the next node in the queue from the updated AST.

Hence, for each removal trial, we record an AST that can be considered as one single data point in form of \texttt{(source, target)} where \texttt{source} is the query node with its features to be extracted and \texttt{target} is the success or failure of our oracle. In total, we collect 30,000 ASTs from which we use 24,000 to train our models and leave aside 6,000 as a test set. 
Our dataset is balanced with 53\% success vs. 47\% failure labels.

\usetikzlibrary{shapes,arrows}
\tikzstyle{block} = [rectangle, draw, fill=blue!10, 
    text width=4em, text centered, minimum height=1em, node distance=2.5cm]
\tikzstyle{block2} = [rectangle, draw, fill=blue!10, 
    text width=8em, text centered, minimum height=1em, node distance=1.5cm]
\tikzstyle{block3} = [rectangle, draw, fill=blue!10, 
    text width=4em, text centered, minimum height=1em, node distance=1.5cm]
 \tikzstyle{line} = [draw, -latex']
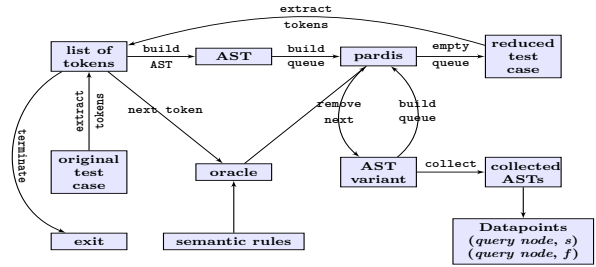
\begin{figure}
\scriptsize
\centering
\resizebox{\columnwidth}{3.5cm}{%
\begin{tikzpicture}[font=\small,every node/.style={font=\bfseries}]
   \node [block] (ast) {AST};
   \node [block, left of=ast] (tokens) {list of tokens};
   \node [block, right of=ast] (pardis) {pardis};
    \node [block, right of=pardis] (reduced) {reduced test case};
    \node [block, below of=tokens] (test) {original test case};
    \node [block, below of=ast] (oracle) {oracle};
    \node [block, below of=pardis] (ASTvariant) {AST variant};
    \node [block, below of=reduced] (logged) {collected ASTs};
    \node [block2, below of=logged] (dps) {Datapoints\\(\textit{query node}, \textit{s}) \\ (\textit{query node}, \textit{f})};
    \node [block3, below of=test] (exit) {exit};
    \node [block2, below of=oracle] (rules) {semantic rules};
    \draw [line] (rules) to (oracle);
    \draw [line] (logged) to (dps);
    \draw [line] (test) to node[midway,sloped,above] {\texttt{extract}}node[midway,sloped,below] {\texttt{tokens}}(tokens);
    \draw [line] (tokens) to node[midway,above] {\texttt{next token}}(oracle);
    \draw [line] (tokens) to [out=215,in=155]node[midway,sloped,above] {\texttt{terminate}}(exit);
    \draw [line] (tokens) to node[midway,above] {\texttt{build}}node[midway,below] {\texttt{AST}}(ast);
    \draw [line] (ast) to node[midway,above] {\texttt{build}}node[midway,below] {\texttt{queue}}(pardis);
    \draw [line] (pardis) to node[midway,above] {\texttt{empty}}node[midway,below] {\texttt{queue}}(reduced);
    \draw [line] (pardis) to [out=225,in=135]node[midway,above] {\texttt{remove}}node[midway,below] {\texttt{next}}(ASTvariant);
    \draw [line] (ASTvariant) to [out=45,in=315] node[midway,above] {\texttt{build}}node[midway,below] {\texttt{queue}}(pardis);
    \draw [line] (oracle) to (pardis);
    \draw [line] (reduced) to [out=160,in=20]node[midway,above] {\texttt{extract}}node[midway,below] {\texttt{tokens}}(tokens);
    \draw [line] (ASTvariant) to node[midway,above] {\texttt{collect}}(logged);
\end{tikzpicture}
}
\caption{Training data collection process. The process terminates when every single token of the test case is either removed or considered to be preserved in the oracle.}
\label{app-fig}
\vspace{-1em}
\end{figure}

\subsection{Feature Extraction}
\label{feat-extract}
We define three different sets of features for query nodes of ASTs collected in \autoref{data-collection}: \textit{query node's type}, \textit{query node's children's types} and \textit{path from the query node to the root of the AST}.
Assume that we want to remove node \texttt{8.BLOCK ITEM} in \autoref{motiv-fig}. This node is our query node with the following features:
\begin{itemize}
    \item Node type: \texttt{BLOCK ITEM}, the type of the node in our C grammar.
    \item Children node types: \texttt{DECLARATION}, \texttt{ASSIGNMENT EXPRESSION}, \texttt{EXPRESSION STATEMENT} and \texttt{JUMP STATEMENT}, types of the immediate children of the node in grammar. 
    \item Path to root: \texttt{BLOCK ITEM}, \texttt{COMPOUND} \texttt{STATEMENT}, \texttt{FUNCTION} \texttt{DEFINITION}, \texttt{COMPILATION} \texttt{UNIT}, types of nodes on the path from the query node to the root of the AST.
\end{itemize}

\textbf{Features Representation.}
For the first feature, we use unique IDs to represent types of query nodes. Each unique ID corresponds to a unique rule type in grammar. We use the C grammar in Perses Normal Form (PNF)~\cite{DBLP:conf/icse/SunLZGS18}. In total, we have 203 grammar rule types and 203 unique IDs corresponding to them. 

We leverage a bag of words model to represent the types of a node's children and the types along the path to the root~\cite{bag-of-words}. For children, we define a bit set of size 204 that represents 203 internal grammar rule types and the keyword \texttt{terminal} used for
leaves of the AST. If a rule type is present among the children's types, we set the corresponding element in the bit set to 1 and 0 otherwise. For the path from the query node to the root, we have a bit set of size 203 (a terminal cannot be on the path to the root).
Similarly, we set the elements of the types of the nodes to 1 if they are present on the path and 0 otherwise.

We feed these features individually or in combination into our training algorithm. The next subsection explains the training process in more detail.

\subsection{Training Models}
\label{train-models}
With Boolean target values of our data points described in \autoref{data-collection}, our problem is a classification problem in which we try to decide which class of removals (valid or invalid) a specific node removal belongs to. To train our models, we use random forests \cite{DBLP:conf/icica/LiuWZ12}, a well-known and efficient classifier that chooses the class with the most votes over all the trees in the forest. 
For each data point in our training set, we feed the features described in \autoref{feat-extract} as source and the result of the node removal (success or failure) as target into our training algorithm.

We build the following five models:
\begin{itemize}
    \item $M_{rf.type}$: model trained using type of the query node as a feature.
    \item $M_{rf.children}$: model trained using types of immediate children of the query node as features.
    \item $M_{rf.path}$: model trained using types of nodes on the path from the query node to the root of the AST as features.
    \item $M_{rf.type.children}$: model trained using types of the query node and its immediate children in combination as features. 
    \item $M_{rf.type.children.path}$: model trained using types of the query node, its children and nodes on the path from the query node to the root all in combination as features. 
\end{itemize}

Note that the process of collecting data, extracting features, and training our models is fairly quick. For instance, it took us 150 minutes in total to collect the GCC torture dataset, extract node types as features, and train the $M_{rf.type}$ model. This is a one-time overhead, and the majority is consumed by the data collection phase. Once data has been collected, multiple models can be trained using the same dataset.
In the following, we explain how we use these models in test case reduction.
\subsection{Querying Models}
\label{model-consult}
We present our approach in \autoref{alg:queue}. Given a test case ($P$) and an oracle function ($\psi$) to reduce the test case, our approach leverages a predictive model ($M$) during test case reduction. The model can be easily integrated into the reduction process.
Similar to Pardis, we traverse a priority queue of syntactically removable nodes. However, for each query node, rather than immediately querying the oracle, we collect its features using \textit{extractFeatures} and query the model instead. The return value of \textit{extractFeatures} depends on the type of model chosen. For instance, it returns the unique ID of the node's type for $M_{rf.type}$, while a bit set is returned for $M_{rf.children}$ and $M_{rf.path}$. A combination of features is returned for $M_{rf.type.children}$ and $M_{rf.type.children.path}$. Each of these models is a Boolean function that receives features of the query node as input and returns a predicted Boolean value.

By querying the model, we can decide whether to continue considering a node. If the model suggests that removing a node will be valid, we proceed to run the oracle and, if the oracle also returns true,
remove the query node and its descendants from the AST because the smaller test preserves the bug.
If the model predicts that removal will be invalid, we take the other branch, add the frontier of syntactically removable descendants to the priority queue similar to Pardis, and continue with attempting to remove the next node in the queue. Note that the key difference between our technique and Pardis is that our approach avoids executing the oracle ($\psi$) when the model predicts that removing the node is invalid. 
\begin{algorithm}\algosize
\scriptsize
\KwIn{$P\in \mathbb{P}$ -- The program (test case) to reduce as an AST}
\KwIn{$\psi: \mathbb{P}\rightarrow\mathbb{B}$ -- Oracle for the property (bug) to preserve where $\mathbb{B}=\{true, false\}$}
\KwIn{$M: \mathbb{S}\rightarrow\mathbb{B}$ -- Model to predict whether a node removal is semantically valid or not. $\mathbb{S}$ is the set of syntactically removable AST nodes represented with their features. }
\KwIn{$Q=\{n_1,n_2,...,n_N\}$ -- Priority queue of syntactically removable AST nodes}
\KwResult{A minimum program $p\in\mathbb{P}$ s.t. $\psi\left(p\right)=true$}
  $p\leftarrow P$
  
  \While{!$Q$.empty()}{
    $node$ $\leftarrow$ $Q$.front()\;
    $features$ = $extractFeatures$(node)\;
    \eIf{$M$(features)}{
      \eIf {$\psi$($p$ - node)}{
        $p$ $\leftarrow$ $p$ $-$ $node$\;
      }
      {
       $Q$.insert(getRemovableFrontier($node$.$children$))\;
      }
    }
    {
     $Q$.insert(getRemovableFrontier($node$.$children$))\;
    }
  }
\Return{p}\;
\caption{Model-guided test case reduction.}
\label{alg:queue}
\end{algorithm}

\section{Results and Discussion}
\label{eval}
We evaluate the performance of our models by answering the following research questions:
\begin{itemize}
    \item \textbf{RQ1.} How do different reducer models perform in terms of reduction time, number of oracle queries (tests) and size of the reduced test case?
    \item \textbf{RQ2.} What are the precision and recall rates of the models?
    \item \textbf{RQ3.} What types of semantic issues (errors and warnings) are predicted correctly by our models  and what types are not? What is the ratio of correct and incorrect predictions with respect to each error and warning type?
\end{itemize}

\subsection{RQ1. Reduction time, number of tests and size}
\label{rq1}
Our tool is available in C++\footnote{https://github.com/golnazgh/Model-Reducer}. We use ANTLR \cite{antlr} to build parse trees (ASTs) from test cases and sklearn-porter\footnote{https://github.com/nok/sklearn-porter} to integrate models into our tool.

\textbf{Experimental Set-up.} As described in \autoref{train-models}, we have five types of models to evaluate: $M_{rf.type}$, $M_{rf.children}$, $M_{rf.path}$, $M_{rf.type.children}$ and $M_{rf.type.children.path}$. Our test data are 14 large test cases with an average size of 70,301 tokens generated by Csmith \cite{DBLP:conf/pldi/YangCER11}, a tool for randomly generating C programs. These test cases, unseen in our training, are reported in Clang and GCC bug repositories and trigger bugs in different versions of these compilers. They have been used to evaluate Pardis and also comprise a large portion of Perses's benchmarks.

In addition to Pardis, the latest state of the art syntax-guided reduction technique, we compare our models with C-Reduce \cite{DBLP:conf/pldi/RegehrCCEEY12}, a tool for automatically reducing C test cases. C-Reduce is a language specific tool with strong reduction power. Our approach is general and can be applied on any input domain.

Results of test case reduction are shown in \autoref{tab::metrics::time}.  Columns with the best results are highlighted. Except for one test case (\texttt{clang-22704}) in which C-Reduce has the best reduction time, our models outperform both Pardis and C-Reduce in terms of time. Our best models in terms of reduction time are $M_{rf.type}$ and $M_{rf.type.children}$ with geomean of 714 and 722 seconds, respectively. Pardis takes 996 seconds on average to reduce these test cases while this number is 1,650 seconds for C-Reduce. This is an improvement of 28\% and 57\% over Pardis and C-Reduce, respectively.
\begin{table*}[t]
\scriptsize
\centering
\caption{Reduction time, number of tests and size of the final reduced test case for different techniques. The best results for each test case are highlighted. For simplification, feature names are used rather than the full model names.}
\label{tab::metrics::time}
\tabcolsep=0.11cm
\begin{adjustbox}{max width=\textwidth}
\begin{tabular}{P{1cm}|P{1.15cm}P{1.15cm}P{1.15cm}P{1.15cm}P{1.15cm}P{1.15cm}P{1.15cm}|P{1.15cm}P{1.15cm}P{1.15cm}P{1.15cm}P{1.15cm}P{1.15cm}P{1.15cm}|P{1.15cm}P{1.15cm}P{1.15cm}P{1.15cm}P{1.15cm}P{1.15cm}P{1.15cm}P{1.15cm}}
\hline
 & \multicolumn{7}{c|}{\thead{Reduction time (sec)}} & \multicolumn{7}{c|}{\thead{Number of tests}} & \multicolumn{8}{c}{\thead{Size (tokens)}}   \\
\cline{2-23}
\thead{Bug} & \thead{type} & \thead{children} & \thead{path} & \thead{type \&\\children} & \thead{type \&\\children\\\& path} & \thead{Pardis} & \thead{C-\\Reduce} & \thead{type} & \thead{children} & \thead{path} & \thead{type \&\\children} & \thead{type \& \\children\\\& path} & \thead{Pardis} & \thead{C-\\Reduce} & \thead{type} & \thead{children} & \thead{path} & \thead{type \&\\children} & \thead{type \& \\children\\\& path} & \thead{Pardis} & \thead{C-\\Reduce} & \thead{Original} \\
\hline
clang-22382 & 7,043 & 7,171 & 7,450 & \cellcolor{blue!25}6,839 & 7,339 & 8,768 & 7,360 & 1,249 & 1,284 & 1,719 & \cellcolor{blue!25}1,214 & 1,566 &  2,011 & 36,637  & 918 & 924 & 1,301 & 1,281 & 1,236 & 354 & \cellcolor{blue!25}80 & 21,068 \\ 
\hline
clang-22704 & 3,244 & 3,256 & 3,141 & 3,236 & 3,126 & 3,226 & \cellcolor{blue!25}1,611 & 3,772 & 3,817 & 3,827 & \cellcolor{blue!25}3,745 & 3,751  & 4,342 & 31,723 & 368 & 355 & 273 & 525 & 449 & 236 & \cellcolor{blue!25}60 & 184,444\\ 
\hline
clang-23309 & \cellcolor{blue!25}656 & 673 & 784 & 739 & 757  & 999 & 2,029 & \cellcolor{blue!25}1,589  & 1,634 & 1,840 & 1,824 & 1,795 &  3,004 & 24,928 &  1,754 & 1,734 & 2,038 & 2,232 & 2,198 & 1,726 & \cellcolor{blue!25}46 & 38,647\\ 
\hline
clang-25900 & \cellcolor{blue!25}398 & 406 & 432 & 406 & 419 & 518 & 802 & 1,077&	1,115&	1,173&	\cellcolor{blue!25}1,073&	1,093 & 1,652 & 20,522 & 771 & 760 & 754 & 823 & 876 & 618 & \cellcolor{blue!25}105 & 78,960\\
\hline
clang-27137 & 2,739 & 2,891 & 3,318 & \cellcolor{blue!25}2,700 & 3,134 & 3,344 & 4,733 &3,059 & 3,314 & 3,964 & \cellcolor{blue!25}2,866 & 3,497& 4,272 & 10,877 & 1,050 & 1,035 & 1,397 & 1,231 & 2,219 &807 & \cellcolor{blue!25}33 & 174,538\\
\hline
clang-27747 & \cellcolor{blue!25}352 & 357 & 402 & 382 & 401 & 449 & 763 & \cellcolor{blue!25}709 & 731 & 879  & 784 & 856 & 1,074 & 16,519 &  424 & 419 & 617 & 497 & 640 & 313 & \cellcolor{blue!25}101 & 173,840\\
\hline
clang-31259 & 435 & 442 & 605 & \cellcolor{blue!25}434 & 582 & 809 & 1,147 & 826 & 850 & 1,061 & \cellcolor{blue!25}822  & 982  &  1,662 & 35,742 & 607 & 602 & 651 & 820 & 764 & 538 & \cellcolor{blue!25}125 & 48,799 \\
\hline
gcc-64990 & 690 & 714 & 807 & \cellcolor{blue!25}687 & 775 & 927 & 2,737 &  1,626 & 1,727 & 2,047 & \cellcolor{blue!25}1,625 & 1,878   & 2,632 & 49,237 & 1,051 & 1,374 & 1,966 & 1,836 & 1,651 &776 & \cellcolor{blue!25}110 & 148,931\\
\hline
gcc-65383  & 530 & 537 & 642 & \cellcolor{blue!25}524 & 625 & 694 & 1,064 & 1,311 & 1,339 & 1,631 & \cellcolor{blue!25}1,290 & 1,625 &  1,839 & 31,738 &  1,010 & 1,004 & 1,693 & 1,277 & 1,636 & 598 & \cellcolor{blue!25}63 & 43,942\\
\hline
gcc-66186 & 355 & 365 & 488 & \cellcolor{blue!25}349 & 440 & 688& 1,381 & 1,008 & 1,038 & 1,349 & \cellcolor{blue!25}961 & 1,222 & 2,562 & 41,364 & 1,276 & 1,271 & 1,639 & 1,623 & 1,667 & 1,176 & \cellcolor{blue!25}142 & 47,481\\
\hline
gcc-66375 & \cellcolor{blue!25}546 & 560 & 726 & 549 & 697 & 1,156 & 1,071 & 1,230 & 1,266 & 1,553 & \cellcolor{blue!25}1,183 & 1,433 & 3,036 & 33,944 & 1,385 & 1,376 & 1,694 & 1,737 & 1,851 & 1,232 & \cellcolor{blue!25}70 & 65,488\\ 
\hline
gcc-70127  & 790 & 802 & 803 & 777 & \cellcolor{blue!25}776 & 1,085 & 3,013 & 1,367 & 1,406 & 1,336 & 1,290 & \cellcolor{blue!25}1,229 &  2,240 & 52,164 &  648 & 641 & 690 & 857 & 822 & 600 & \cellcolor{blue!25}71 & 154,816\\
\hline
gcc-70586 & \cellcolor{blue!25}1,207 & 1,223 & 1,371 & \cellcolor{blue!25}1,207 & 1,437 & 1,709 & 2,915 & 2,036 & 2,073 & 2,334 & \cellcolor{blue!25}1,918 & 2,405  &  3,491 & 21,269 & 1,613 & 1,608 & 1,681 & 2,090 & 1,936 & 1,497 & \cellcolor{blue!25}101 & 212,259\\ 
\hline
gcc-71626  & \cellcolor{blue!25}53 & 54 & \cellcolor{blue!25}53 & 54 & \cellcolor{blue!25}53 & 56 & 366 & 216 & 227 & 206 & 221 & \cellcolor{blue!25}197  &  264 & 11,144 &  65 & 64 & 59 & 64 & 59 & 58 & \cellcolor{blue!25}47 & 6,133\\ 
\hline
geomean & \cellcolor{blue!25}714 & 729 & 826 & 722 & 803 & 996 & 1,650 &  1,251 & 1,294 & 1,477 & \cellcolor{blue!25}1,245 & 1,395 & 2,065 & 26,895 & 743 & 750 & 889 & 947 & 993 & 574 & \cellcolor{blue!25}76 & 70,301\\
\hline
median & \cellcolor{blue!25}601 & 617 & 755 & 618 & 727 & 963 & 1,496 & 1,280 & 1,312 & 1,592 & \cellcolor{blue!25}1,252 & 1,500 & 2,401 & 31,731& 964 & 964 & 1,349 & 1,254 & 1,436& 609 & \cellcolor{blue!25}76 & 72,224\\ 
\hline
\end{tabular}
\end{adjustbox}
\end{table*}

In addition, our models have smaller number of tests compared to Pardis and C-Reduce. As described in \autoref{model-consult}, we proceed with performing a test (querying the oracle) only if the model predicts a semantically valid removal. This generates a smaller number of tests for our models compared to existing techniques. The improvement in reduction time we obtain by skipping tests confirms the more expensive process of querying oracles compared to the models. 

Our models maintain similar reduction power with slight increase in the final reduced size compared to Pardis with $M_{rf.type}$, our best model in terms of reduction time generating reduced test cases of size 743 tokens on average versus an average of 574 tokens for Pardis. One reason for the increase is the possibility of different local minima \cite{DBLP:journals/tse/ZellerH02} when different reduction techniques may follow different paths during reduction and each generate a correct but different reduced test case. Another likely reason is because of how our approach works. We skip tests if our model predicts that they will be semantically invalid. However, we \emph{may} skip a valid test due to a wrong prediction that may generate larger reduced test cases. We discuss this more in \autoref{prec-recall}.

In comparison with C-Reduce, our approach generates larger reduced test cases (743 vs. 76 tokens on average). This is expected since C-Reduce is a language specific tool highly customized for effectively reducing C test cases. Our approach is general and does not require specific knowledge of a given language. However, our models can be utilized as both individual and complimentary to the state of the art techniques. In the following, we combine $M_{rf.type}$, our best model in terms of reduction time with C-Reduce and Pardis to measure the reduction performance.

\noindent\textbf{$\boldsymbol M_{\boldsymbol{rf.type}}$ $\boldsymbol\rightarrow$ C-Reduce.} In order to measure the overall performance of test case reduction when $M_{rf.type}$ is used as a preprocessing step with another technique, we feed reduced test cases already generated by $M_{rf.type}$ to C-Reduce as inputs to be reduced again.
Results of this study are shown in \autoref{tab::metrics::combine::creduce}. For each test case, the better value of reduction time, size and number of tests between C-Reduce and $M_{rf.type}$ $\rightarrow$ C-Reduce is highlighted. For individual techniques, $M_{rf.type}$ and C-Reduce, the reduction time is the time it took each approach to reduce the original test case (also shown in \autoref{tab::metrics::time}). For $M_{rf.type}$ $\rightarrow$ C-Reduce, the reduction time is the time it took $M_{rf.type}$ to reduce the original test case added by the time it took C-Reduce to reduce the test case already reduced by $M_{rf.type}$. The size columns show the size of the final reduced test case.
As can be seen, in 11 out of 14 test cases, within shorter reduction time, we are able to generate a reduced test case with a size almost the same as C-Reduce using $M_{rf.type}$ $\rightarrow$ C-Reduce. 
On average, $M_{rf.type}$ $\rightarrow$ C-Reduce reduces test cases to 82 tokens versus 76 tokens by C-Reduce while having a shorter reduction time (1,278 vs. 1,650 seconds). In addition, we are able to reduce the total number of tests by 40\% on average when using $M_{rf.type}$ $\rightarrow$ C-Reduce compared to C-Reduce.

\begin{table*}[!th]
\scriptsize
\centering
\caption{Performance metrics using $M_{rf.type}$ as a preprocessing step for C-Reduce and Pardis. Better values between each state of the art technique and its combined form are highlighted for each metric.}
\label{tab::metrics::combine::creduce}
\tabcolsep=0.11cm
\begin{adjustbox}{max width=\textwidth}
\begin{tabular}{|P{1.5cm}|P{1.1cm}||P{1.3cm}|P{1.4cm}||P{1.1cm}|P{1.2cm}||P{1.1cm}||P{1.3cm}|P{1.4cm}||P{1.1cm}|P{1.2cm}||P{1.1cm}||P{1.3cm}|P{1.4cm}||P{1.1cm}|P{1.2cm}|}
\hline
& \multicolumn{5}{c|}{\thead{Reduction time (sec)}} & \multicolumn{5}{c|}{\thead{Size (tokens)}} &
\multicolumn{5}{c|}{\thead{Number of tests}} 
\\
\cline{2-16}
\thead{Bug} & \thead{$\boldsymbol M_{\boldsymbol{rf.type}}$} & \thead{C-Reduce} & \textbf{$\boldsymbol M_{\boldsymbol{rf.type}}$ $\boldsymbol\rightarrow$ C-Reduce} & \thead{Pardis} & \textbf{$\boldsymbol M_{\boldsymbol{rf.type}}$ $\boldsymbol\rightarrow$ Pardis} & \thead{$\boldsymbol M_{\boldsymbol{rf.type}}$} & \thead{C-Reduce} & \textbf{$\boldsymbol M_{\boldsymbol{rf.type}}$ $\boldsymbol\rightarrow$ C-Reduce} &  \thead{Pardis} & \textbf{$\boldsymbol M_{\boldsymbol{rf.type}}$ $\boldsymbol\rightarrow$ Pardis} & \thead{$\boldsymbol M_{\boldsymbol{rf.type}}$} & \thead{C-Reduce} & \textbf{$\boldsymbol M_{\boldsymbol{rf.type}}$ $\boldsymbol\rightarrow$ C-Reduce} &  \thead{Pardis} & \textbf{$\boldsymbol M_{\boldsymbol{rf.type}}$ $\boldsymbol\rightarrow$ Pardis}  \\
\hline
clang-22382 & 7,043 & \cellcolor{blue!25}7,360 & 10,291 & 8,768 & \cellcolor{blue!25}8,050 & 918 & \cellcolor{blue!25}80 &  82 & \cellcolor{blue!25}354 & \cellcolor{blue!25}354 & 1,249 & 36,637 & \cellcolor{blue!25}17,832 &\cellcolor{blue!25}2,011 & 2,260 \\ 
\hline
clang-22704 & 3,244 & \cellcolor{blue!25}1,611 & 3,460 & \cellcolor{blue!25}3,226 & 3,284 & 368 & 60 & \cellcolor{blue!25}57 & \cellcolor{blue!25}236  & \cellcolor{blue!25}236 & 3,772& 31,723 & \cellcolor{blue!25}10,517 & 4,342&  \cellcolor{blue!25}4,226  \\
\hline
clang-23309  & 656 & 2,029  & \cellcolor{blue!25}1,559 & 999 & \cellcolor{blue!25}890 & 1,754 & \cellcolor{blue!25}46 & 125 & 1,726  & \cellcolor{blue!25}1,419 & 1,589 & \cellcolor{blue!25}24,928& 30,105 & \cellcolor{blue!25}3,004& 3,599 \\ 
\hline
clang-25900  & 398 & 802 &  \cellcolor{blue!25}801 & 518 & \cellcolor{blue!25}484 & 771 & \cellcolor{blue!25}105 &  106 & 618 & \cellcolor{blue!25}608 & 1,077 & 20,522& \cellcolor{blue!25}18,967 &  \cellcolor{blue!25}1,652 & 1,889 \\
\hline
clang-27137  & 2,739  & 4,733 &  \cellcolor{blue!25}3,028 & 3,344 & \cellcolor{blue!25}3,145 & 1,050 & \cellcolor{blue!25}33 & 39 & 807 & \cellcolor{blue!25}710 & 3,059 & 10,877&  \cellcolor{blue!25}8,775 & \cellcolor{blue!25}4,272& 4,804 \\ 
\hline
clang-27747  & 352 & 763 &  \cellcolor{blue!25}603 & 449 & \cellcolor{blue!25}414 & 424 & 101 & \cellcolor{blue!25}69 & 313 & \cellcolor{blue!25}311  & 709 & 16,519& \cellcolor{blue!25}10,677 &  \cellcolor{blue!25}1,074 & 1,232  \\ 
\hline
clang-31259  & 435 & 1,147 & \cellcolor{blue!25}964 & 809 & \cellcolor{blue!25}720 & 607 & \cellcolor{blue!25}125 & 137& \cellcolor{blue!25}538 & 541  & 826 &35,742& \cellcolor{blue!25}21,778 & \cellcolor{blue!25}1,662 & 1,736   \\
\hline
gcc-64990 & 690 & 2,737 &  \cellcolor{blue!25}1,331 & 927  & \cellcolor{blue!25}843 & 1,051 & \cellcolor{blue!25}110 & 119 & \cellcolor{blue!25}776 & 918 & 1,626& 49,237 & \cellcolor{blue!25}24,700 & \cellcolor{blue!25}2,632 & 2,950 \\ 
\hline
gcc-65383   & 530 & 1,064 &  \cellcolor{blue!25}946 & 694 & \cellcolor{blue!25}608 & 1,010 & 63 & \cellcolor{blue!25}50 & 598 & \cellcolor{blue!25}475 & 1,311 & 31,738& \cellcolor{blue!25}14,474 & \cellcolor{blue!25}1,839& 2,200 \\ 
\hline
gcc-66186 & 355 & 1,381 & \cellcolor{blue!25}1,057 & 688 & \cellcolor{blue!25}606 & 1,276 & 142 & \cellcolor{blue!25}140 & \cellcolor{blue!25}1,176 & 1,178 & 1,008& 41,364 & \cellcolor{blue!25}25,064 & \cellcolor{blue!25} 2,562 &  3,144  \\
\hline
gcc-66375  & 546 & \cellcolor{blue!25}1,071 & 1,202 & 1,156 & \cellcolor{blue!25}793   & 1,385 & 70 & \cellcolor{blue!25}66 & \cellcolor{blue!25}1,232 & 1,291 & 1,230 & 33,944&  \cellcolor{blue!25}24,901 & \cellcolor{blue!25}3,036 & 3,178  \\ 
\hline
gcc-70127 & 790  & 3,013  & \cellcolor{blue!25}1,212 & 1,085  & \cellcolor{blue!25}932 & 648 & \cellcolor{blue!25}71 & 98 & 600 & \cellcolor{blue!25}598  & 1,367 & 52,164 & \cellcolor{blue!25}17,013 &  \cellcolor{blue!25}2,240 & 2,376  \\
\hline
gcc-70586  &  1,207 & 2,915 & \cellcolor{blue!25}2,022  & 1,709  & \cellcolor{blue!25}1,530 & 1,613 & \cellcolor{blue!25}101 &  102 & \cellcolor{blue!25}1,497 & 1,500  & 2,036 & \cellcolor{blue!25}21,269 & 30,345 & \cellcolor{blue!25}3,491 & 4,384  \\ 
\hline
gcc-71626  & 53 & 366 & \cellcolor{blue!25}101 & 56 & \cellcolor{blue!25}55 & 65 & \cellcolor{blue!25}47 & \cellcolor{blue!25}47 & \cellcolor{blue!25}58 & \cellcolor{blue!25}58 & 216 & 11,144& \cellcolor{blue!25}3,617 & \cellcolor{blue!25}264 & 283 \\  
\hline
geomean  & 714 & 1,650  & \cellcolor{blue!25}1,278 & 996 & \cellcolor{blue!25}893 & 743 & \cellcolor{blue!25}76 & 82 & 574& \cellcolor{blue!25}560 & 1,251 & 26,895 & \cellcolor{blue!25}16,269 & \cellcolor{blue!25}2,065 & 2,316 \\ 
\hline
median  & 601 & 1,496 & \cellcolor{blue!25}1,207 & 963 & \cellcolor{blue!25}818 & 964 & \cellcolor{blue!25}76 &  90& 609 & \cellcolor{blue!25}603 & 1,280 & 31,731 & \cellcolor{blue!25}18,400 & \cellcolor{blue!25}2,401 & 2,663 \\ 
\hline
\end{tabular}
\end{adjustbox}
\end{table*}

\noindent\textbf{$\boldsymbol M_{\boldsymbol{rf.type}}$ $\boldsymbol\rightarrow$ Pardis.} Similar to $M_{rf.type}$ $\rightarrow$ C-Reduce, we feed test cases reduced by $M_{rf.type}$ to Pardis to gain more reduction if possible. In 13 out of 14 cases, we obtain test cases with size similar to Pardis while performing faster. 
$M_{rf.type}$ $\rightarrow$ Pardis generates reduced test cases with 560 tokens on average compared to Pardis with 574 tokens while performing reduction in 893 seconds compared to Pardis with 996 seconds reduction time on average.

\begin{figure*}[h]
\scriptsize
\begin{subfigure}{.33\linewidth}
\centering
\scriptsize
\includegraphics[scale=0.27]{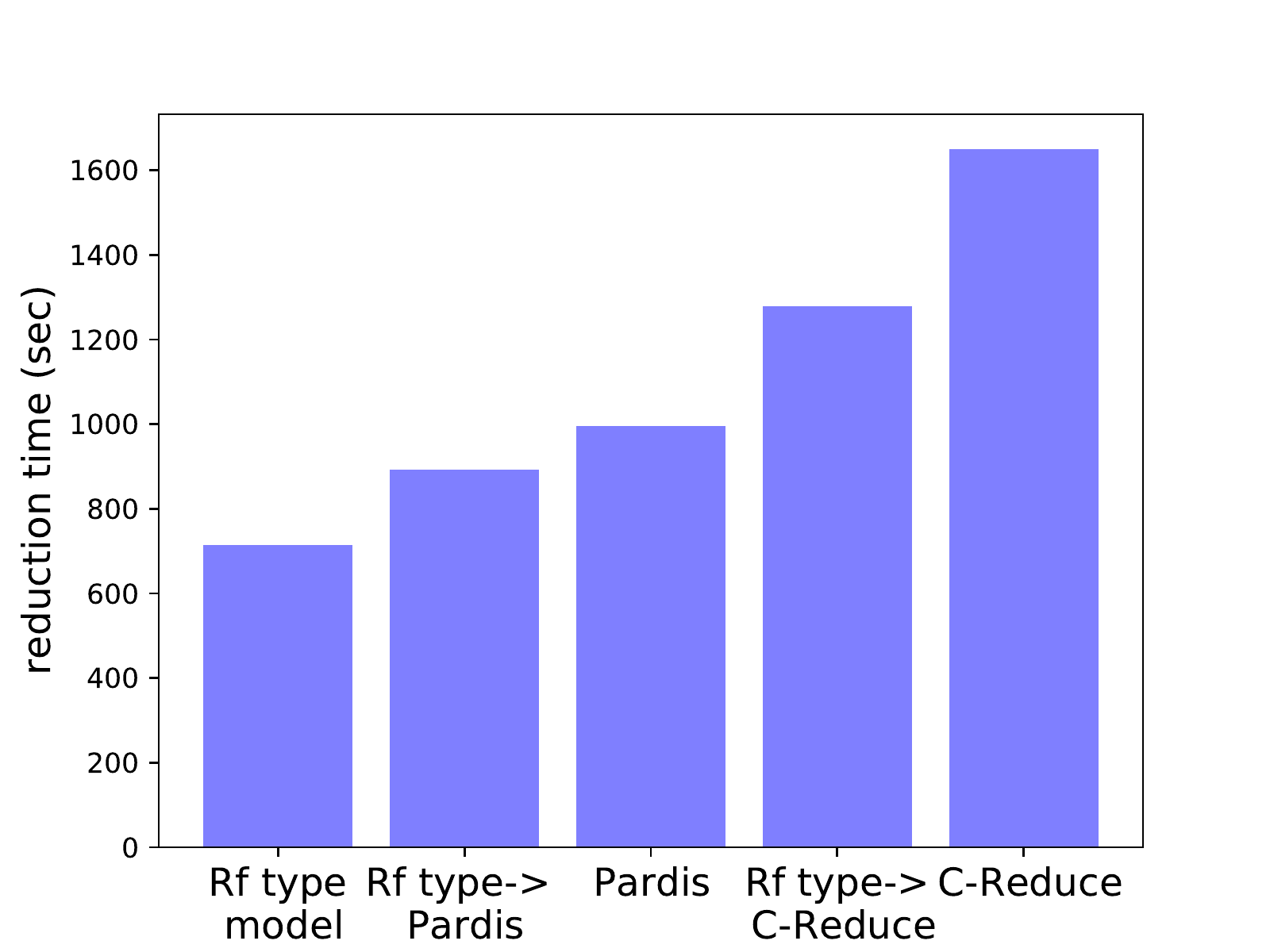}
\caption{reduction time}
\label{fig:sub1}
\end{subfigure}%
\begin{subfigure}{.33\linewidth}
\centering
\scriptsize
\includegraphics[scale=0.27]{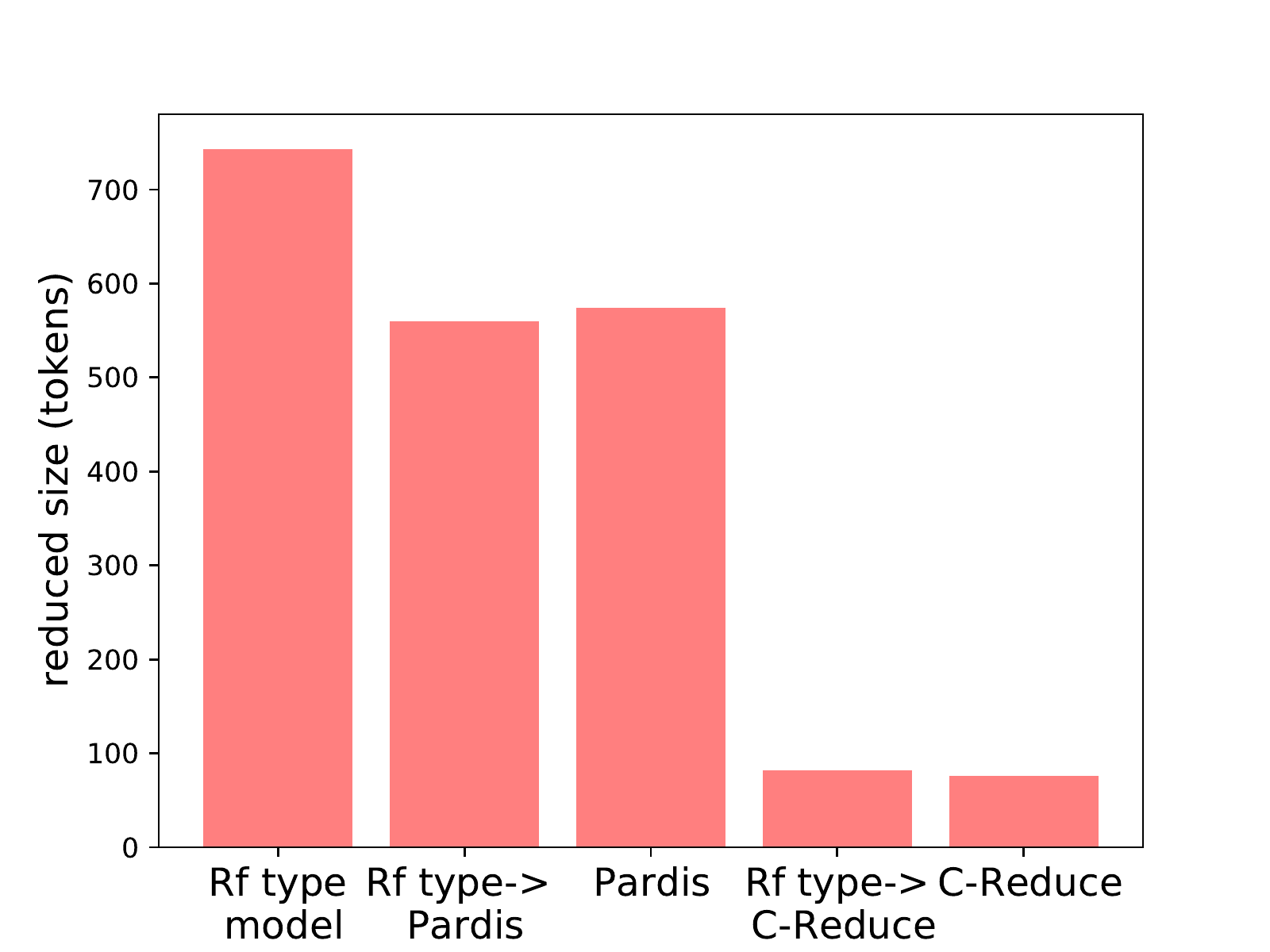}
\caption{final reduced size}
\label{fig:sub2}
\end{subfigure}
\begin{subfigure}{.33\linewidth}
\centering
\scriptsize
\includegraphics[scale=0.27]{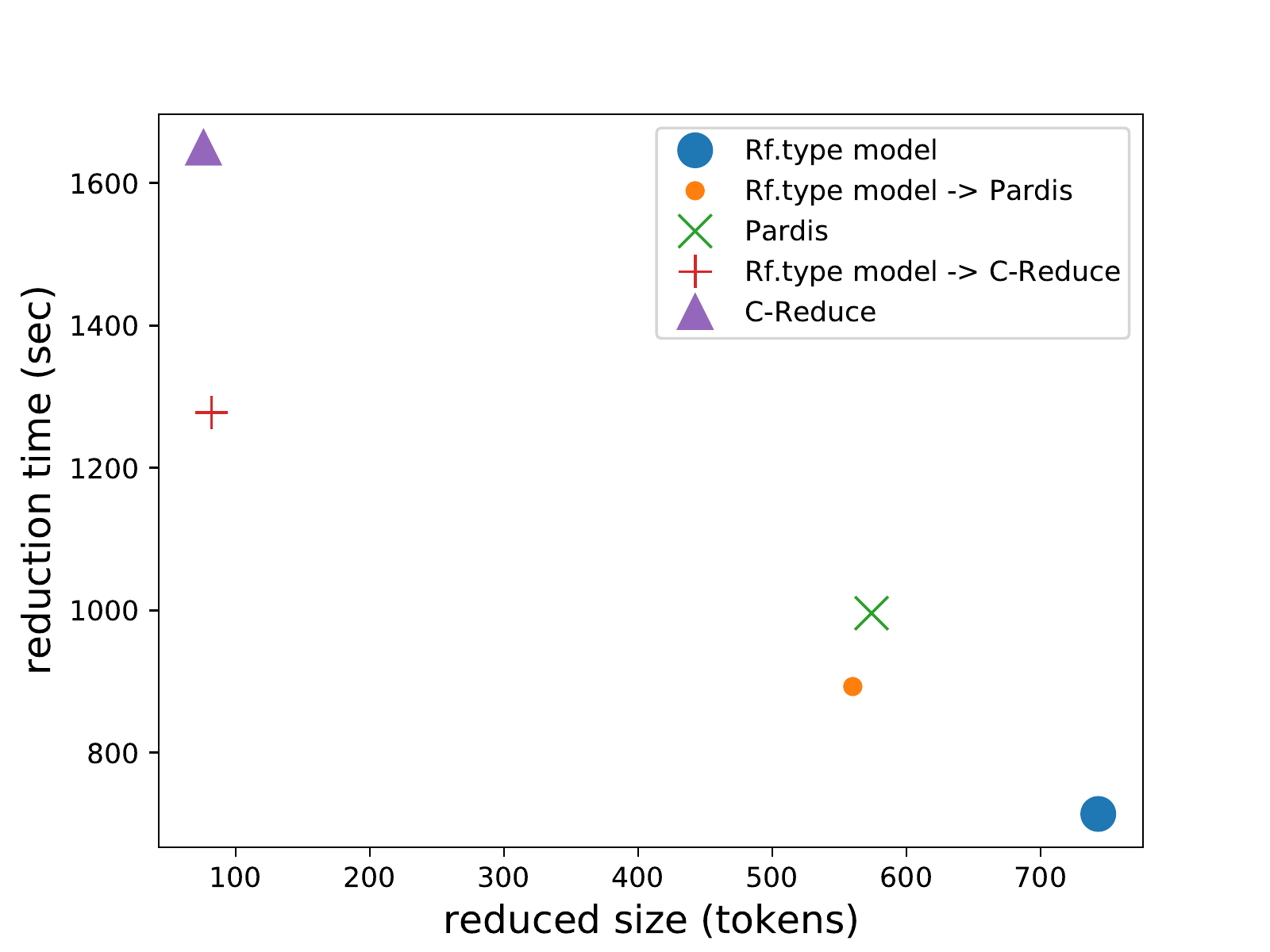}
\caption{precision vs. efficiency}
\label{fig:sub3}
\end{subfigure}
\caption{Geomean of reduction time and reduced size for individual and hybrid techniques.}
\label{fig:geomean-comparison}
\end{figure*}

\textbf{Assessing trade offs.} Measuring the performance of $M_{rf.type}$ $\rightarrow$ C-Reduce and $M_{rf.type}$ $\rightarrow$ Pardis shows us a trade-off with respect to the test case reduction time and size of the reduced test case for a user (debugger). If a slight increase in size is acceptable, $M_{rf.type}$ is the fastest of the models examined and also generates test cases of acceptable size. However, if size is more important than time for the user, $M_{rf.type}$ $\rightarrow$ C-Reduce or $M_{rf.type}$ $\rightarrow$ Pardis are still faster than C-Reduce and Pardis and can generate reduced test cases of the same size. \autoref{fig:geomean-comparison} depicts an overview of reduction time and size for different techniques and how they stand with respect to each other. The fastest approach is $M_{rf.type}$ with largest reduced size whereas C-Reduce takes the longest to generate the smallest test cases. Our hybrid techniques, $M_{rf.type}$ $\rightarrow$ C-Reduce and $M_{rf.type}$ $\rightarrow$ Pardis,  improve the reduction time of the state of the art techniques while preserving the same reduction power.

\subsection{RQ2. Precision and recall rates}
\label{prec-recall}
Here, we discuss the possible categories of our models' outcomes (predicted values) with respect to the oracle's outcome (actual value) and compute the precision and recall rates of our models.
\tikzstyle{circle} = [circle]
\tikzstyle{decision} = [diamond, draw, fill=blue!20, 
    text width=4em, text badly centered, node distance=1cm, inner sep=0pt]
\tikzstyle{block} = [rectangle, draw, fill=blue!20, 
    text width=5em, text centered, rounded corners, minimum height=4.5em, node distance=3cm]
\tikzstyle{block2} = [rectangle, draw, fill=blue!20, 
    text width=6.5em, text centered, rounded corners, minimum height=4.5em, node distance=1cm]
 \tikzstyle{line} = [draw, -latex']
\begin{figure}
\centering
\resizebox{0.8\columnwidth}{4cm}{%
\begin{tikzpicture}
[font=\small, edge from parent,
    every node/.style={top color=white, bottom color=blue!25, 
    rectangle,rounded corners, minimum size=6mm, draw=blue!75,
    very thick, drop shadow, align=center},
    edge from parent/.style={draw=blue!50,thick},
    level 1/.style={sibling distance=3cm},
    level 2/.style={sibling distance=7cm}, 
    level 3/.style={sibling distance=3cm}, 
    level distance=2.5cm,
]
    \node [decision] (init) {Model's outcome?}
    child { node [block] (TN) {\textbf{false $\boldsymbol\rightarrow$ TN}\\ Semantically \textbf{invalid} tests \textbf{filtered} by model}
    }
    child { node [block] (TP) {\textbf{true $\boldsymbol\rightarrow$ TP}\\ Semantically \textbf{valid} tests \textbf{not filtered} by model}
         child{ node [decision] (evaluate) {Oracle's\\ outcome?}
              child {node[block2] {\textbf{true $\boldsymbol\rightarrow$ $\boldsymbol{TP_{op}}$} \\ Oracle \\passed}}
              child {node[block2] {\textbf{false $\boldsymbol\rightarrow$ $\boldsymbol{TP_{ns}}$} \\ Non-semantic \\oracle failure} 
              }
         }
    }
     child { node [block] (FP) {\textbf{true $\boldsymbol\rightarrow$ FP}\\ Semantically \textbf{invalid} tests falsely \textbf{not filtered} by model}}
    child{ node [block] (FN) {\textbf{false $\boldsymbol\rightarrow$ FN}\\ Semantically \textbf{valid} tests falsely \textbf{filtered} by model}
     child{ node [decision] (evaluate) {Oracle's\\ outcome?}
              child {node[block2] {\textbf{true $\boldsymbol\rightarrow$ $\boldsymbol{FN_{op}}$} \\ Oracle\\ passed}}
              child {node[block2] {\textbf{false $\boldsymbol\rightarrow$ $\boldsymbol{FN_{ns}}$} \\ Non-semantic \\oracle failure} 
              }
         }
    };

\end{tikzpicture}
}
\caption{A breakdown of reducer model's outcomes.}
\vspace{-1.5em}
\label{model-outcome-fig}
\end{figure}
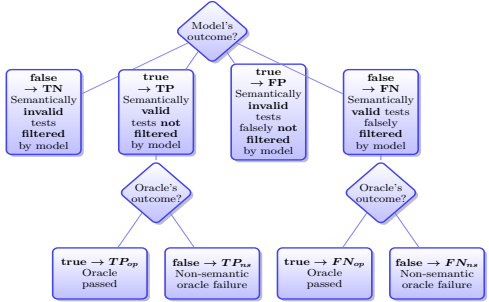

\noindent\textbf{Model outcomes.} With respect to the outcome of an oracle query, the predicted outcomes of our models belong to one of the following categories (\autoref{model-outcome-fig}):    
\begin{itemize}
    \item \textit{True positive (TP):} Model correctly predicts that removal of the query node is semantically valid. After passing the semantics validity check, these tests will be subdivided into two subcategories:
    \begin{itemize}
        \item \textit {Oracle pass ($TP_{op}$):} The return value of the oracle query is true for these tests leading to a successful removal and reducing the test case.
        \item \textit {Non-semantic failure ($TP_{ns}$):} These tests fail to remove the node from the AST of the test case. However, their failure is caused by reasons other than semantic invalidity. They may fail well-definedness checks ( \autoref{motiv}) or not preserve the property of interest (not exhibit the bug).
    \end{itemize}
    \item \textit{True negative (TN):} Model correctly predicts that removal of the node is semantically invalid. 
    \item \textit{False positive (FP):} Model falsely predicts that removal of the query node is semantically valid.
    \item \textit{False negative (FN):} Model falsely predicts that removal of the query node is semantically invalid. Using our approach, these tests will not be executed due to a false prediction. If tests were to be executed, they would have the subcategories of oracle pass ($FN_{op}$) and non-semantic failure ($FN_{ns}$) similar to the true positives.
\end{itemize}
\subsubsection{Distribution of models' outcome categories}
\label{distribution-study}
We conduct a study to compute the distribution of our models' outcome categories when reducing 13 test cases from the same benchmark we evaluated our models on in \autoref{rq1}. We exclude the smallest test case, \texttt{gcc-71626} since there are no blacklisted warnings or well-definedness checks included in its oracle  obtained from Perses and Pardis studies.

We reduce test cases using our three main models, $M_{rf.type}$, $M_{rf.children}$ and $M_{rf.path}$ and record the model's prediction of the semantic validity along with the actual outcome of the oracle for each test. We define three actual outcomes for an oracle: 
\begin{itemize}
    \item \textit{Passed:} Test was successful. Oracle passed and node got removed.
    \item \textit{Semantic checks failed:} Test was unsuccessful. Oracle failed due to the semantic invalidity of the test case variant and node did not get removed.
    \item \textit{Well-definedness or property checks failed:} Test was unsuccessful. Semantic checks passed but oracle failed due to failing well-definedness checks or not preserving the bug.
\end{itemize}

Note that although we guide the reduction process by our models' suggestions for removal of each node, in this study we query the oracle every time we try to remove a node no matter what the prediction of the model is. This is how we can have pairs of values with predicted and actual outcomes used to compute the categories of \textit{TNs}, \textit{TPs}, \textit{FPs} and \textit{FNs}. \begin{figure*}
\scriptsize
\begin{subfigure}{.33\linewidth}
\centering
\includegraphics[scale=0.3]{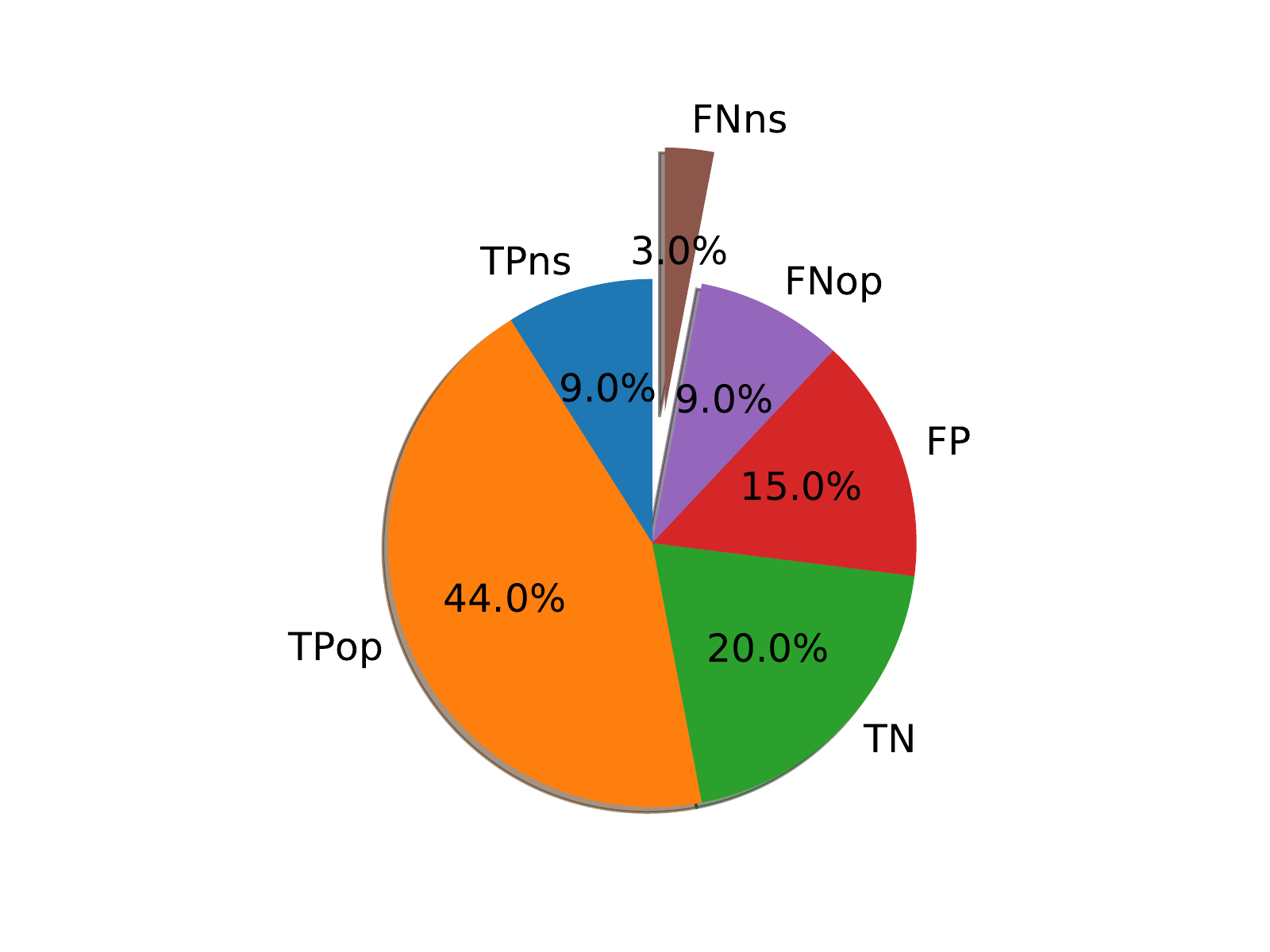}
\caption{$M_{rf.type}$}
\label{fig:sub1}
\end{subfigure}%
\begin{subfigure}{.33\linewidth}
\centering
\includegraphics[scale=0.3]{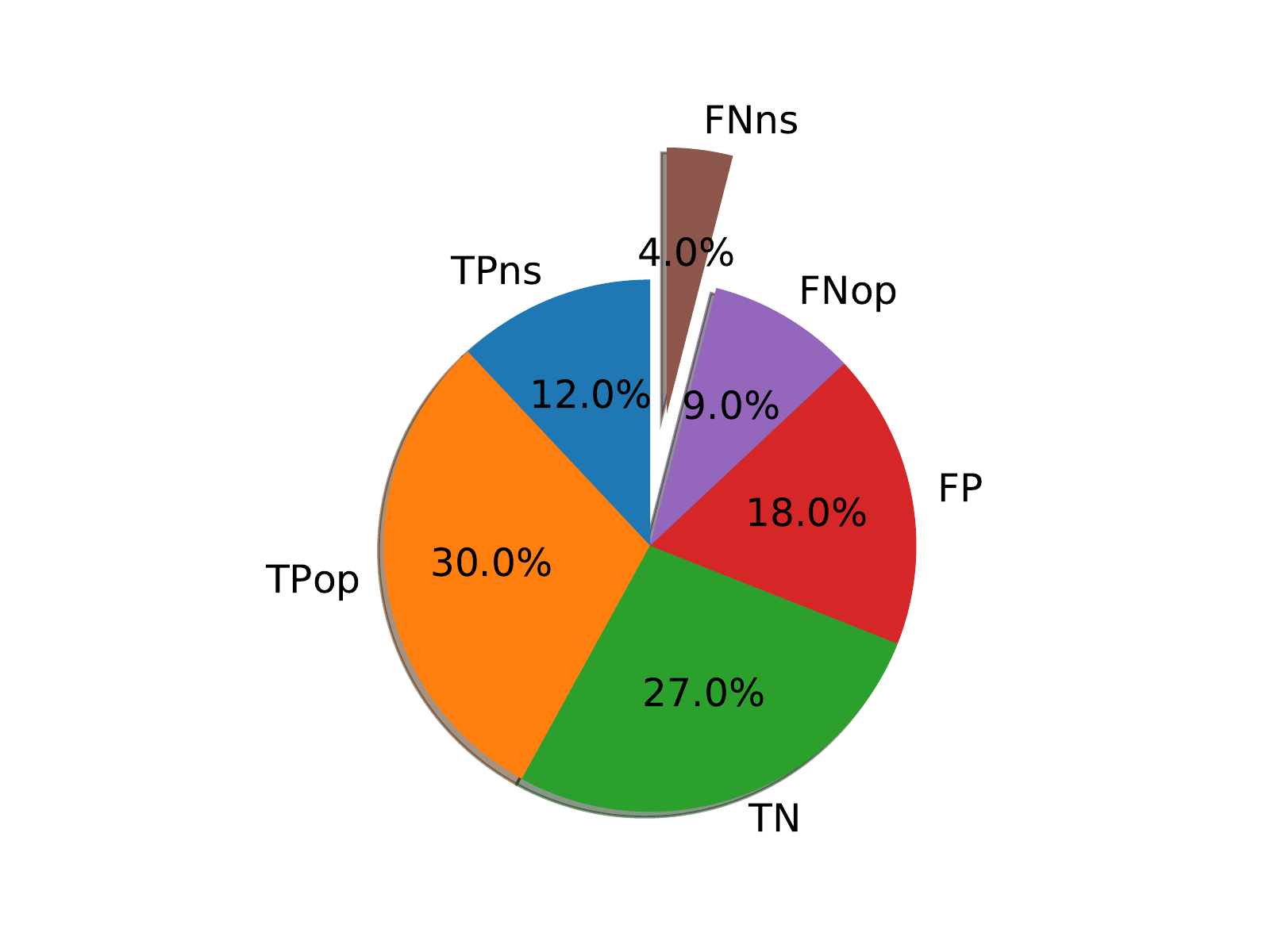}
\caption{$M_{rf.children}$}
\label{fig:sub2}
\end{subfigure}
\begin{subfigure}{0.33\linewidth}
\centering
\includegraphics[scale=0.3]{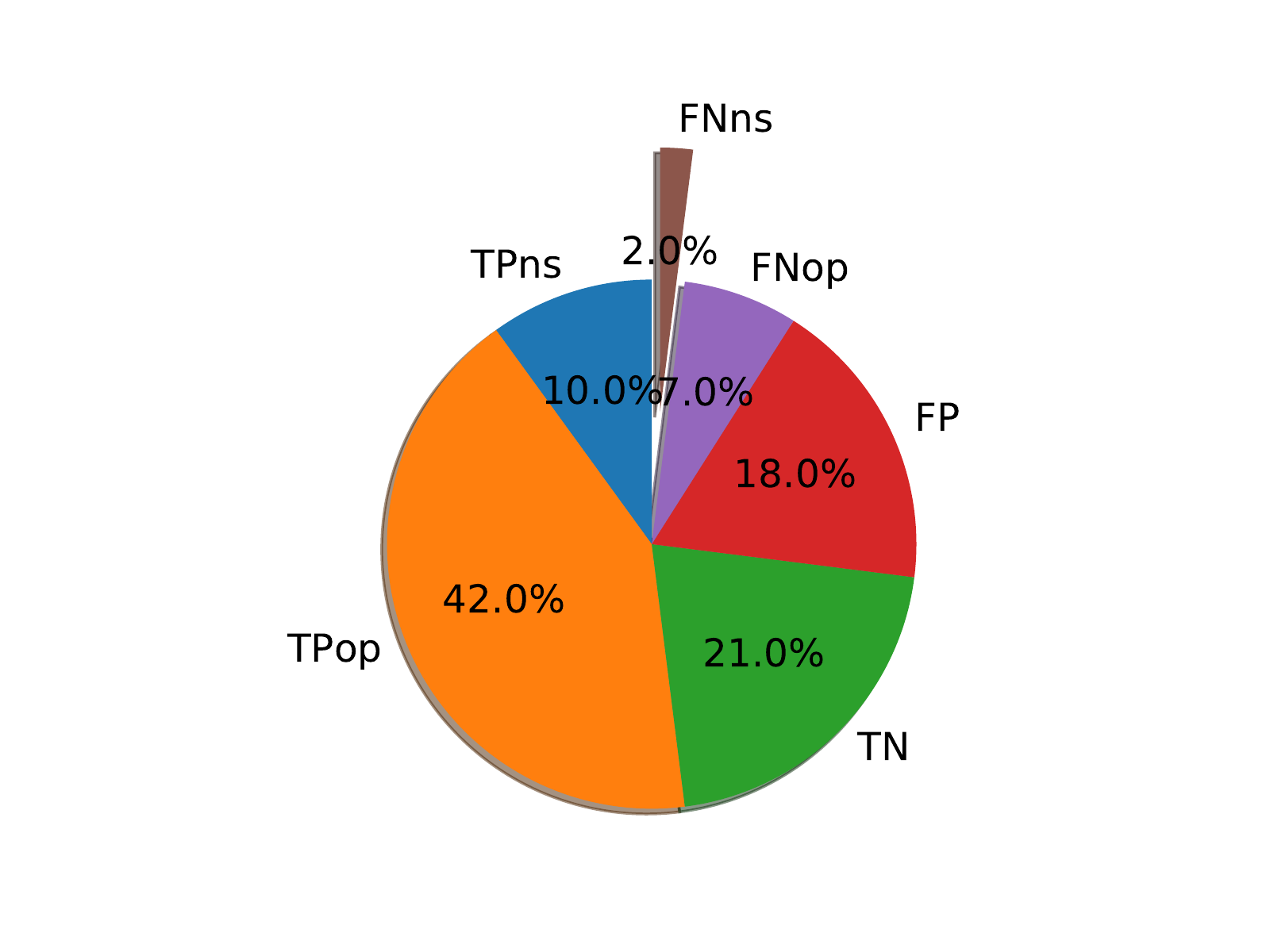}
\caption{$M_{rf.path}$}
\label{fig:sub3}
\end{subfigure}
\caption{Distribution of model's outcome categories for our three main models.}
\label{fig:piechart::color}
\end{figure*}

\autoref{fig:piechart::color} shows the distribution of our three models' outcome categories. We collected pairs of predicted (by model) and actual (by oracle) outcome values for 37,021, 37,183 and 41,697 total tests performed when reducing 13 test cases querying the three models, $M_{rf.type}$, $M_{rf.children}$ and $M_{rf.path}$, respectively. As can be seen, $M_{rf.type}$ correctly filters 20\% of all the tests, identifying semantically invalid inputs (true negatives), and it correctly allows the execution of 53\% of all tests (true positives), from which 44\% lead to an oracle pass and 9\% fail for reasons other than semantic invalidity.  27\% of predictions by $M_{rf.type}$ are incorrect, which is discussed more in \autoref{rq3}. $M_{rf.children}$ and $M_{rf.path}$ have similar distributions with 69\% and 73\% true predictions, respectively.  

\subsubsection{Precision and recall rates (real test set)}
\label{dynamic-test}
With true positives, false positives and false negatives collected in our distribution study, we compute the precision and recall rates of our models. The results are shown in \autoref{tab::metrics::precision::actual}. Since these values are computed on our fuzzer-generated test cases with real bugs, we call them precision and recall rates on the \textit{real test set}.
On average, $M_{rf.type}$ has the highest precision rate of 77\% whereas  $M_{rf.path}$ has the highest recall rate with an average of 84\%. 

\begin{table}[!th]
\scriptsize
\centering
\caption{Precision and recall rates of models on real test set.}
\label{tab::metrics::precision::actual}
\tabcolsep=0.11cm
\begin{adjustbox}{max width=\textwidth}
\begin{tabular}{|P{1.5cm}|P{0.5cm}|P{0.65cm}|P{0.5cm}|P{0.5cm}|P{0.5cm}|P{0.65cm}|P{0.5cm}|P{0.5cm}|}
\hline
& \multicolumn{4}{c|}{\thead{Precision (\%)}} &  \multicolumn{4}{c|}{\thead{Recall (\%)}}
\\
\cline{2-9}
\thead{Model\\\textcolor{white}{ids}} & \thead{min} & \thead{mean} & \thead{med} & \thead{max}  & \thead{min} & \thead{mean} &\thead{ med} & \thead{max}  \\
\hline
$M_{rf.type}$ & 65 & 77 & 78 & 81 & 68 & 80 &  81 & 84\\ 
\hline
$M_{rf.children}$ & 69 & 70  & 70 &  75 & 76 & 77& 77 & 79\\ 
\hline
$M_{rf.path}$  & 70 & 74 & 74 & 76 & 80 & 84 & 84 & 85\\ 
\hline
\end{tabular}
\end{adjustbox}
\end{table}
\subsubsection{Precision and recall rates (synthetic test sets)}
In addition to computing precision and recall rates using test cases with real bugs, we compute these values for our GCC torture test set with 6,000 ASTs. Since these test cases do not have a real bug and we synthesize their oracles by including the semantic validity and token preservation criteria (\autoref{data-collection}), we refer to them as \textit{synthetic test cases}. We obtained 75\%, 76\% and 72\% as the precision rates of $M_{rf.type}$, $M_{rf.children}$ and $M_{rf.path}$, respectively. With respect to the recall rates, $M_{rf.path}$ had the highest value with 84\% followed by $M_{rf.type}$ and $M_{rf.children}$ with 78\% and 61\%, respectively.

\subsection{RQ3. Prediction of semantic invalidity types}
\label{rq3}
\begin{figure}
\scriptsize
\centering
\includegraphics[scale=0.36]{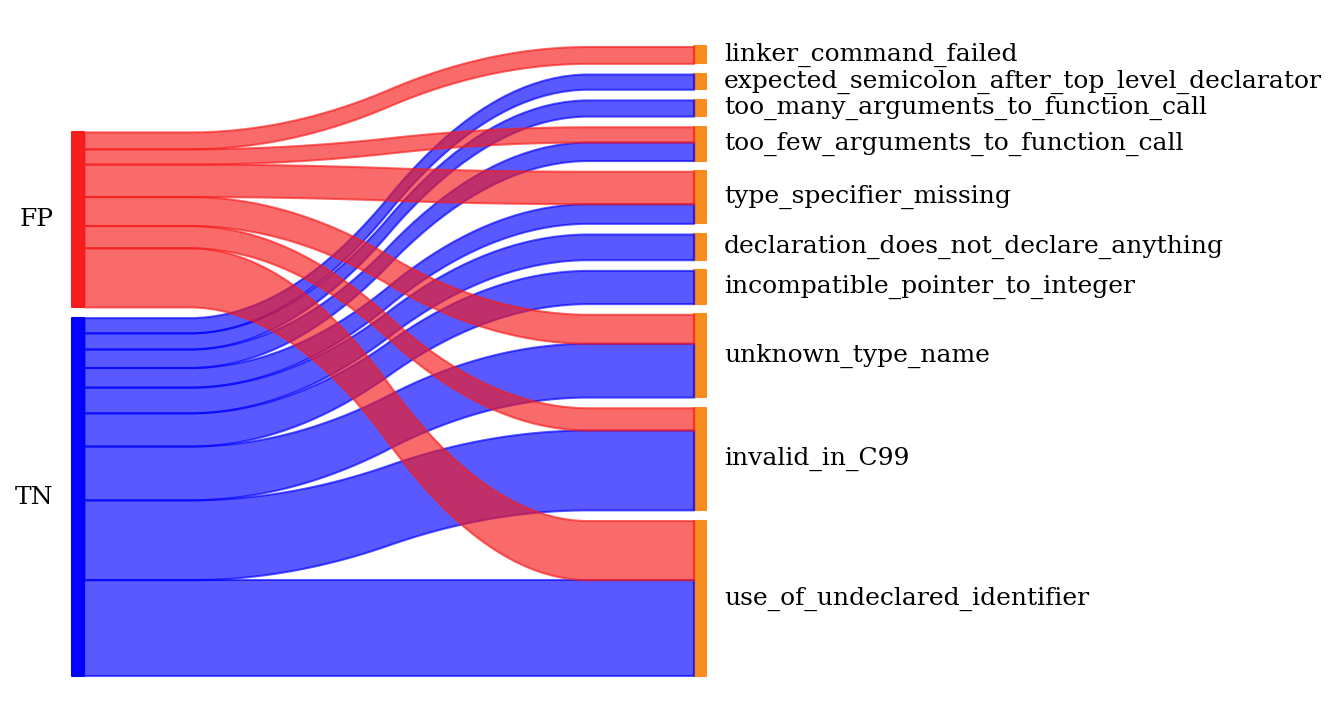}
 \caption{Issues filtered (TN) and missed (FP) by $M_{rf.type}$.}
\label{TN-FP-breakdown}
\end{figure}

Finally, we measure the performance of our models with respect to correct and incorrect prediction of different types of semantic issues. More precisely, we examine which error and blacklisted warning types are predicted correctly by our models and which ones are not. To this end, we investigate two sets of models' outcomes: true negatives that are the errors and warnings filtered by our models and false positives that are errors and warnings missed by our models. \autoref{TN-FP-breakdown} depicts the breakdown of true negatives (TN) and false positives (FP) for $M_{rf.type}$.

\texttt{Use of undeclared identifier} is one of the common errors during reduction in which a declaration is removed before its use in the program. Our model correctly filters most of these errors while still missing a portion of them. In \autoref{discuss}, we discuss possible improvements of the models to capture more of these errors in the future. \texttt{Invalid in C99} is another major category of semantic issues detected by our model. There are issues such as \texttt{declaration does not declare anything}, \texttt{too} \texttt{many} \texttt{arguments} \texttt{to} \texttt{function} \texttt{call} and \texttt{expected semicolon after top level declarator} entirely filtered by our model.

\section{Improvements and Future Work}
\label{discuss}
We used a specific set of features to train our models. However, these features are not the only properties to reason about the validity of the tests. We may require more complex features to improve precision of our models. We observed that \texttt{use of undeclared identifier} is a common error type. As a future direction, we can define features that capture dependencies between declarations and uses of entities more precisely. For instance, a path from a declaration to a use or some dependency pairs representing these entities. However, a possible challenge is increased noise in the training data that may adversely affect the performance of the models.

Another possibility is to train individual models for individual types of semantic issues. This may improve the precision of each model on a specific type of issue, however since the type of the error or warning is unknown at the time of the removal, the type of the model to query will also be unknown. Some solutions such as querying all models or a few of them at each removal can mitigate the problem. However, it is unclear how efficient these solutions will be.

Our approach is general and can be applied on any input with a grammar. In this work, we focused on C test cases to have a fair comparison with Pardis and C-Reduce. However, models can be trained and used on other domains. In addition, our approach is highly adaptable and can be integrated easily with  models trained using different learning techniques. Hence, it can directly benefit from possibly more precise classifiers in future.

\section{Related Work}
We discuss related work in program reduction and use of machine learning in programming languages.

\noindent\textbf{Program (Test Case) Reduction.}
There is a wide variety of techniques used to reduce test cases and inputs in general. Delta Debugging \cite{DBLP:journals/tse/ZellerH02}, its improvements \cite{DBLP:conf/icsoft/Hodovan016, DBLP:conf/qrs/GharachorluS18}, Hierarchical Delta Debugging   \cite{DBLP:conf/icse/Misherghi06} and its variants \cite{hdd-thesis, DBLP:conf/sigsoft/0001HG18, DBLP:conf/icsm/HodovanKG17, DBLP:conf/icse/HodovanKG17} are among the most famous ones. Despite the generality and simplicity of these techniques, they suffer from a considerably large number of syntactically and semantically invalid tests.     To reduce the number of these invalid tests, modernized HDD  \cite{DBLP:conf/sigsoft/Hodovan016} takes advantage of a context free grammar to minimize syntactically invalid tests. A similar and more recent work is Perses \cite{DBLP:conf/icse/SunLZGS18} that leverages a priority queue driven approach. Although Perses speeds up test case reduction by filtering syntactically invalid tests, it suffers from a concept called \textit{priority inversion} in which a long time of reduction is spent on portions of the AST with the least benefit. Pardis \cite{DBLP:conf/fase/GharachorluS19} mitigates this problem and outperforms Perses by ordering the queue such that portions of the AST with high value are reached during the first phases of the reduction. However, these techniques lack the ability of reasoning about the semantic validity of a test. Our work is the first one that tries to determine the removability of a node using both syntactic and semantic properties.

Among other related work, C-Reduce \cite{DBLP:conf/pldi/RegehrCCEEY12} is a language specific tool suitable for reducing C test cases. Our approach is general and can be applied to any input with a given grammar. GTR \cite{DBLP:conf/kbse/HerfertPP17} applies transformations to the AST using a generalized tree reduction algorithm. Despite the generality, semantic validity remains neglected in this work.

\noindent\textbf{Machine learning in programming languages.}
Models to perform tasks related to programs are continuously trained and used. Here, we mention a few of them: Decision trees to propose probabilistic model for code completion or repair \cite{DBLP:conf/oopsla/RaychevBV16}, neural machine translation to repair the code that does not compile \cite{DBLP:conf/sigsoft/0001RJGA19} and reinforcement learning to debloat programs while preserving specifications of the program's functionality \cite{DBLP:conf/ccs/HeoLPN18}.
Finally, there are works using machine learning to help facilitate static analysis \cite{mlForUnsoundSA, mlForStaticTypestate, mlForIncrementalSA, DBLP:journals/pacmpl/ChaeOHY17}. Our approach is the first one to leverage models to predict semantic validity of a program.

\section{Conclusion}
A bug report with a smaller test case is more likely to be investigated in software issue trackers. The lack of an automated test case reduction technique capable of reducing test cases while trying to preserve both syntactic and semantic validity is what we addressed in this paper and proposed an approach to mitigate it. Using random forests, we trained models by extracting multiple syntactic features from a large corpus of test cases. Our models, easily integrated into a reducer, skip a test if it is predicted as semantically invalid. Compared to the state of the art technique, our best model with an average precision of 77\% speeds up test case reduction by almost 30\% on a suite of large fuzzer-generated test cases used in bug reports of two well-known C compilers.


\section*{Acknowledgment}
This research was supported by the Natural Sciences and Engineering Research Council of Canada.

\newpage
\bibliographystyle{IEEEtran}
\bibliography{references}

\end{document}